\documentclass[showpacs,amsmath,amssymb]{revtex4}

\usepackage{graphicx}
\usepackage{dcolumn}
\usepackage{bm}
\usepackage{epsfig}

\begin{document}

\def\0#1#2{\frac{#1}{#2}}
\def\bct{\begin{center}} \def\ect{\end{center}}
\def\beq{\begin{equation}} \def\eeq{\end{equation}}
\def\bea{\begin{eqnarray}} \def\eea{\end{eqnarray}}
\def\nnu{\nonumber}
\def\n{\noindent} \def\pl{\partial}
\def\g{\gamma}  \def\O{\Omega} \def\e{\varepsilon} \def\o{\omega}
\def\s{\sigma}  \def\b{\beta} \def\p{\psi} \def\r{\rho}
\def\G{\Gamma} \def\S{\Sigma} \def\l{\lambda}

\title{Expanding the thermodynamical potential and the analysis of the possible phase diagram of deconfinement in FL model}
\author{Song~Shu}
\affiliation{Faculty of Physics and Electronic Technology, Hubei
University, Wuhan 430062, China}
\author{Jia-Rong~Li}
\affiliation{Institute of Particle Physics, Hua-Zhong Normal
University, Wuhan 430079, China}
\begin{abstract}
The deconfinement phase transition is studied in the FL model at
finite temperature and chemical potential. At MFT approximation,
the phase transition can only be the first order in the whole
$\mu-T$ phase plane. By a Landau expansion we further study the
phase transition order and the possible phase diagram of
deconfinement. We discuss the possibilities of second order phase
transitions in FL model. By our analysis the cubic term in the
Landau expansion could be cancelled by the high order
fluctuations. By an ansatz of the Landau parameters, we obtain the
possible phase diagram with both first and second order phase
transition including the tricritical point which is similar to
that of the chiral phase transition.
\end{abstract}  \maketitle

\section{Introduction}
It is generally believed that at sufficiently high temperatures
and densities there is a QCD phase transition from normal nuclear
matter to QGP~\cite{ref1,ref2}. Theoretically there are two kinds
of phase transitions associated with different symmetries for two
opposite quark mass limit. For $N_f=2+1$ massless quark flavors,
the QCD lagrangian posses a chiral symmetry $SU(N_f)_R\times
SU(N_f)_L$, which is associated with the chiral phase transition.
In the heavy quark limit, QCD reduces to a pure $SU(N_c)$ gauge
theory which is invariant under a global $Z(N_c)$ center symmetry.
This symmetry is associated with the deconfinement phase
transition. The orders of these phase transitions have been
studied extensively~\cite{ref3,ref4,ref5} and still remained to be
an interesting problem~\cite{ref6a,ref6b,ref6c,ref6d}. For chiral
phase transition at finite temperature in the chiral limit, the
quark-antiquark condensate $\langle\bar q_Rq_L\rangle$ serves as a
good order parameter. The order of the phase transition depends on
the quark flavors. For $N_f=3$ massless quark flavors, it is a
first order phase transition. For $N_f=2$ massless quark flavors,
it is a second order phase transition. At finite densities, the
chiral phase transition have been studied by many effective
models~\cite{ref7,ref8,ref9}. It is generally regarded that at
high densities it is a first order phase transition. In the
$\mu-T$ phase diagram, from first chiral phase transition to
second order phase transition there exists a tri-critical
point(TCP). For deconfinement phase transition, it has not a good
order parameter except for infinite quark mass limit, at which the
Polyakov loop severs as an order parameter~\cite{ref10,ref11}. In
recent studies the Polyakov loop has been combined into the chiral
models,such as Nambu-Jona-Lasinio model~\cite{ref12,ref13} and
linear sigma model~\cite{ref14,ref15,ref16}, which allows to
investigate the deconfinement phase transition within the chiral
models. Though the Polyakov loop is not a good order parameter, it
still serves as an indicator of a rapid crossover towards
deconfinement. As we know in the Landau theory, for the study of
the phase transition and the transition order, one should find a
good order parameter. Once it is identified, the thermodynamic
functions could be expanded over this order parameter and the
transition order could be well studied. For the deconfinement
phase transition, besides the Polyakov loop, one can also search
for other proper order parameters in the effective field models.
In the earlier studies of deconfinement, the bag models had been
often used to investigate the confinement mechanics and the
thermodynamics of deconfinement phase transition. In this paper we
wish to use the effective bag model to study the deconfinement
phase transition and mainly focus on the study of the transition
order and the possible phase diagram of the deconfinement,
especially the possible influence on the phase diagram by the
fluctuations.

The model we used here is Friedberg-Lee(FL) soliton bag model. The
FL model has been widely discussed in past
decays~\cite{ref17,ref18,ref19}. It has been very successful in
describing phenomenologically the static properties of hadrons and
their behaviors at low energy. The model consists of quark fields
interacting with a phenomenological scalar field $\s$. The $\s$
field is introduced to describe the complicated nonperturbative
features of QCD vacuum. It naturally gives a color confinement
mechanism in QCD theory. The model has been also extended to
finite temperatures and densities to study deconfinement phase
transition~\cite{ref20,ref21,ref22,ref23,ref24}. Here we will try
to identify the proper order parameter in this model and make an
analysis of deconfinement phase transition.

The organization of this paper is as follows: in section 2 we give
a brief introduction of the FL model. The thermodynamic potential
is derived and deconfinement phase transition is discussed at
finite temperatures and densities at mean field theory (MFT)
approximation. In section 3, we make a Landau expansion of the
thermodynamic potential. In this way the transition order is
studied by analyzing the Landau coefficients. By an ansatz of
Landau coefficients we discuss the possible phase diagram of
deconfinement in FL model. The last section is the summary.

\section{The thermodynamic potential and deconfinement phase transition in FL model at MFT}
We start from the Lagrangian of the FL model, \bea {\cal
L}=\bar\psi(i\gamma_\mu\pl^\mu-g\s)\psi+\012(\pl_\mu\s)(\pl^\mu\s)-U(\s),
\eea where\bea U(\s)=\01{2!}a\s^2+\01{3!}b\s^3+\01{4!}c\s^4+B.
\eea $\p$ represents the quark field, and $\s$ denotes the
phenomenological scalar field. $a, b, c, g$ and $B$ are the
constants which are generally fitted in with producing the
properties of hadrons appropriately at zero temperature. We shift
the $\s$ field as $\s\rightarrow\bar\s+\s'$ where $\bar\s$ and
$\s'$ are the vacuum expectation value and the fluctuation of the
$\s$ field respectively. Then the lagrangian becomes \bea {\cal
L}_{eff}=\bar\psi(i\gamma_\mu\pl^\mu-m_q)\psi+\012(\pl_\mu\s')(\pl^\mu\s')-\012m_\s^2\s'^2-U(\bar\s),
\eea where \bea
U(\bar\s)=\01{2!}a\bar\s^2+\01{3!}b\bar\s^3+\01{4!}c\bar\s^4+B.
\eea $m_q=g\bar\s$ and $m_\s^2=a+b\bar\s+\012c\bar\s^2$ are the
effective masses of the quark and $\s$ fields respectively. The
interactions associated with the fluctuation $\s'$, such as
$\s'^3$, $\s'^4$ and $\bar\psi\s'\psi$, are neglected in MFT
approximation.

According to finite temperature field theory, the partition
function is \bea
Z=\int[d\bar\p][d\p][d\s']\exp\left[\int_0^{\beta}d\tau\int d^3
{\bf x}({\cal L}_{eff} +\mu\psi^{\dag}\psi)\right].  \eea where
$\mu$ is chemical potential of quarks. Completing the integration
in partition function $Z$, together with the thermodynamic
potential: $\O=-TlnZ$, at mean field level, we could obtain \bea
\O=U(\bar\s)+\01{\b}\int \0{d^3\bf p}{(2\pi)^3}\ln (1-e^{-\b
E_{\s}})-\0{\g}{\b}\int \0{d^3\bf p}{(2\pi)^3}\left[\ln
(1+e^{-\b(E_q-\mu)}) + \ln (1+e^{-\b(E_q+\mu)})\right],
\label{mean} \eea where $\b$ is the inverse of the temperature $T$
and $\g$ is a degenerate factor that $\g=2(spin)\times
2(flavor)\times 3(color)$. In addition, $E_\s=\sqrt{\vec
p^2+m_\s^2}$ and $E_q=\sqrt{\vec p^2+m_q^2}$.
\begin{figure}[tbh]
\begin{center}
\includegraphics[width=210pt,height=150pt]{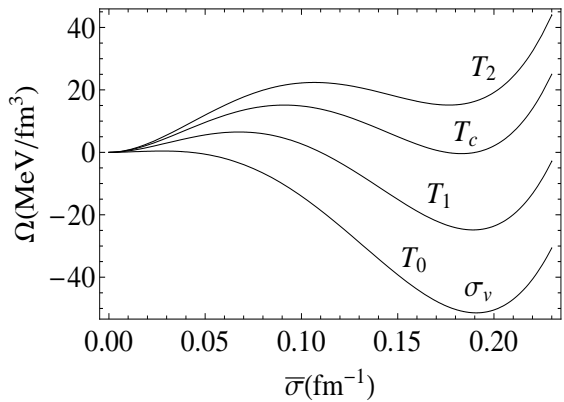}
\end{center}
\caption{The thermodynamical potentials for different temperatures
and zero chemical potential: $T_0=0MeV$, $T_1=100MeV$,
$T_{c}=121MeV$ and $T_2=130MeV$.}\label{f1}
\end{figure}

\begin{figure}[tbh]
\begin{center}
\includegraphics[width=210pt,height=150pt]{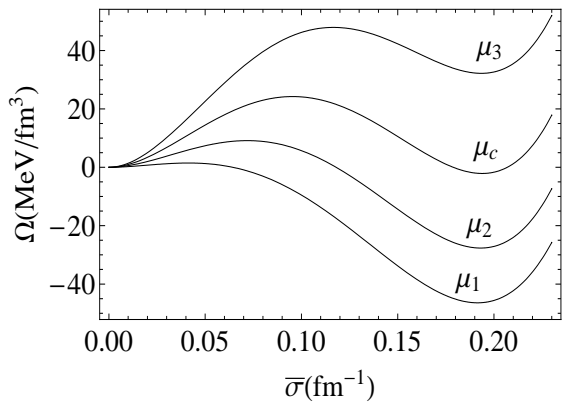}
\end{center}
\caption{The thermodynamical potentials for different chemical
potentials and fixed temperature at $T=50MeV$: $\mu_1=100MeV$,
$\mu_2=200MeV$, $\mu_{c}=255MeV$, and $\mu_3=300MeV$.}\label{f2}
\end{figure}

In our calculation, the parameters are chosen to be
$a=17.7fm^{-2}, b=-1457.4fm^{-1}, c=20000, g=12.16$. The effective
mass of $\s$ field is fixed at $m_\s=550MeV$~\cite{ref22}. Then
one could plot $\O$ versus $\bar\s$ for different $T$ as shown in
Fig.\ref{f1}. At zero temperature, where $\O=U(\bar\s)$, there are
two minima of the thermodynamic potential: one corresponds to the
perturbative vacuum at $\bar\s=0$, another corresponds to the
physical vacuum at $\bar\s=\s_v$. The system is stabled at the
physical vacuum at $\bar\s=\s_v$. It is well known that at this
time the quarks are confined in a soliton bag, and the system is
in a hadronic phase. With temperature increased, the physical
vacuum $\bar\s=\s_v$ is lifted up, while the quarks has been still
confined until the two vacuums degenerate. At this time the
deconfinement phase transition occurs, and the phase transition
temperature is $T=T_c$. After that, the system is stabled at the
perturbative vacuum $\bar\s=0$, where the quarks are deconfined
and the system is in a deconfined phase. This is a first order
phase transition.

One can also plot the $\O$ versus $\bar\s$ at different $\mu$ for
$T=50MeV$ as shown in Fig.\ref{f2}. The deconfinement phase
transition takes place at $\mu=\mu_c$ where the two vacuums
degenerate. The analysis of deconfinement phase transition at
finite chemical potential is similar to that at finite
temperature.
\begin{figure}[tbh]
\begin{center}
\includegraphics[width=210pt,height=150pt]{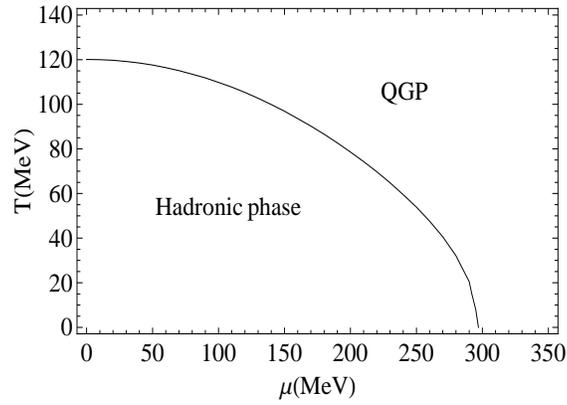}
\end{center}
\caption{The $\mu-T$ phase diagram of deconfinement at MFT in the
FL model.}\label{f3}
\end{figure}

One can obtain the $\mu-T$ phase diagram as shown in Fig.\ref{f3}.
In the whole $\mu-T$ phase plane, the transition is first order.

\section{A Landau expansion and the possible phase diagram of deconfinement phase transition}
\begin{figure}[tbh]
\begin{center}
\includegraphics[width=210pt,height=150pt]{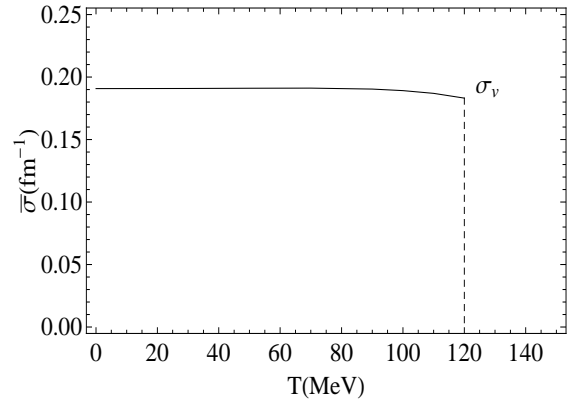}
\end{center}
\caption{$\bar\s$ as a function of $T$ at zero chemical potential
in the FL model.}\label{f4}
\end{figure}

In above discussion, we know at MFT approximation in FL model the
deconfinement phase transition is first order. One can plot the
$\bar\s$ as a function of $T$, as shown in Fig.\ref{f4}. It could
be seen that at $T=T_c$, $\bar\s$ jumps from nonzero value
$\bar\s=\s_v$ to zero value $\bar\s=0$. In confined phase
$\bar\s\neq 0$; in deconfined phase $\bar\s=0$. Here $\bar\s$
could be viewed as an order parameter of deconfinement phase
transition in FL model, so we can do a Landau expansion of $\O$
based on $\bar\s$ and make a thorough investigation of the phase
transition order.

At the MFT approximation, from equation (\ref{mean}) the
thermodynamic potential could be power expanded by $\bar\s$ with
$\bar\s^2$, $\bar\s^3$ and $\bar\s^4$. However, the analytical
forms of the coefficients of the expansion are difficult to be
obtained. Here we will write down the effective form of the
expansion as \bea
\O=\012A(T,\mu)\bar\s^2+\01{3!}B(T,\mu)\bar\s^3+\01{4!}C(T,\mu)\bar\s^4,
\label{exp} \eea where $A(T,\mu)$, $B(T,\mu)$ and $C(T,\mu)$ are
the effective parameters which could be determined by a numerical
fitting process. That means at certain $T$ and $\mu$ from the
configuration of the $\O$ versus $\bar\s$ one could fit the curve
by the $\bar\s^2$, $\bar\s^3$ and $\bar\s^4$ to obtain the values
of $A(T,\mu)$, $B(T,\mu)$ and $C(T,\mu)$. By the equation
(\ref{exp}), from Landau theory, it is clear that the cubic term
$\bar\s^3$ plays crucial role in determination of the transition
order. At MFT approximation, the fitting results indicate that
$B(T,\mu)$, as a negative value, will keep decreasing with
temperature and/or chemical potential increasing. That means this
term will never be zero, therefore the transition order of
deconfinement at MFT approximation can only be first order.

Now we suppose equation (\ref{exp}) is the general form of
expansion of thermodynamical potential by order parameter $\bar\s$
in FL model. And we regard the corrections coming from the
fluctuations will effectively modify the parameters $A(T,\mu)$,
$B(T,\mu)$ and $C(T,\mu)$. In principle they could be calculated
by self-consistently resumming the higher order loop diagrams led
by the fluctuations of $\s'$. However it is very difficult to
evaluate these corrections in this way. In the following we will
treat the coefficients $A(T,\mu)$, $B(T,\mu)$ and $C(T,\mu)$ as
the free parameters and make a general study of the phase
transition order on the FL model by Landau theory.
\begin{figure}[tbh]
\begin{center}
\includegraphics[width=210pt,height=150pt]{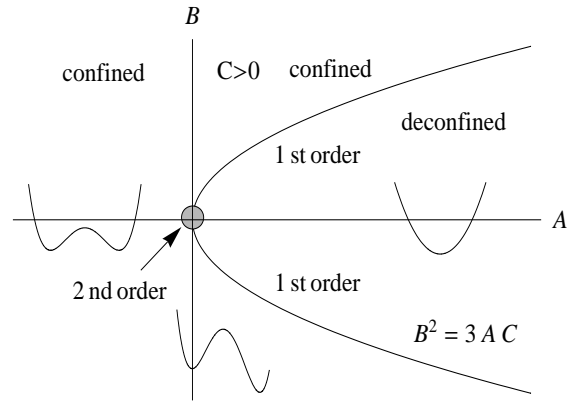}
\end{center}
\caption{Phase diagram of deconfinement on the $A-B$ plane in the
FL model.}\label{f5}
\end{figure}

In Landau theory, one can make a derivative of the thermodynamic
potential to $\bar\s$ as \bea
\0{d\O}{d\bar\s}=A(T,\mu)\bar\s+\012B(T,\mu)\bar\s^2+\01{3!}C(T,\mu)\bar\s^3=0.
\eea One can obtain three solutions:\bea \bar\s_{1}=0, \ \ \ \ \
\bar\s_{2,3}=\0{-3B\pm\sqrt{9B^2-24AC}}{2C}. \eea In our case, we
assume $C>0$ which guarantees that the vacuums are the minima.
When $3B^2\le8AC$, there is only one minimum at $\bar\s=0$. When
$3B^2>8AC$, there are two minima. They correspond to the
perturbative vacuum at $\bar\s=0$ and the physical vacuum at
$\bar\s=\s_v$. When the two minima degenerate, one can obtain the
condition that: $B^2=3AC$, at which the deconfinement phase
transition takes place. Thus one can draw the critical line of the
deconfinement phase transition in the plane of $B$ versus $A$ as
shown in Fig.\ref{f5}. The phase plane has been divided into two
parts: the left area beside the line in the plane represents the
confined phase, while the right area the deconfined phase. By
analyzing the variation of the vacuum, one can obtain that the
deconfinement phase transition can be either first or second
order. If the system goes across the critical line at $B\ne 0$,
the transition is first order. If the system goes across the line
at $B=0$, the transition is second order.
\begin{figure}[tbh]
\begin{center}
\includegraphics[width=210pt,height=150pt]{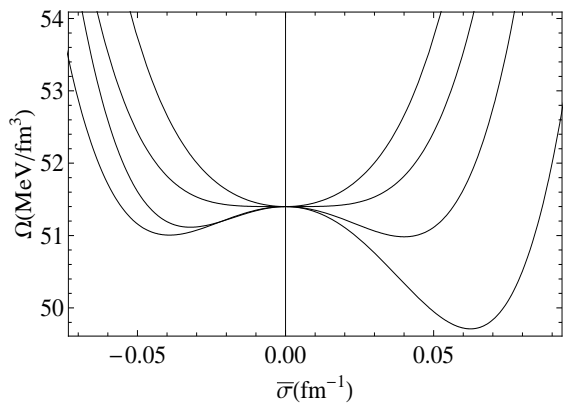}
\end{center}
\caption{The thermodynamical potentials for different temperatures
and zero chemical potential. The temperatures are $100MeV, 150MeV,
180MeV$ and $200MeV$ from bottom to top.}\label{f6}
\end{figure}

From above discussion by Landau theory, we know there may be a
second order phase transition in FL model, while at MFT level, the
deconfinement phase transition can only be first order. But if we
consider fluctuations beyond MFT, there are maybe additional terms
which cancel the cubic $\bar\s^3$ term. The second order phase
transition may be possible. That means the parameter $B(T,\mu)$
will go to zero before the transition takes place. The system will
evolve from left area to right area across the critical line by
the axis origin in the Fig.\ref{f5}. In our former calculation at
MFT, the fluctuations of $\s'$ in the Lagrangian have been
neglected. These terms are possibly important in the cancellation
of the cubic term. However, it is very difficult to calculate the
thermodynamic potential including these fluctuations from the
Lagrangian in FL model. In the following, we will make an ansatz
based on the form of the Landau expansion of the thermodynamic
potential to mimic the deconfinement phase transition which have
both first and second order phase transition.
\begin{figure}[tbh]
\begin{center}
\includegraphics[width=210pt,height=150pt]{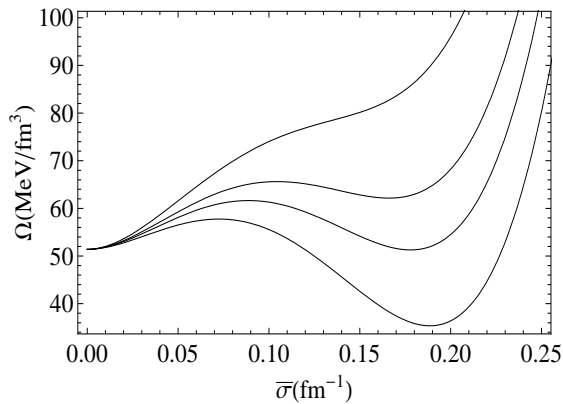}
\end{center}
\caption{The thermodynamical potentials for different chemical
potentials and zero temperature. The chemical potentials are
$350MeV, 392MeV, 420MeV$ and $470MeV$ from bottom to
top.}\label{f7}
\end{figure}

We can devise a possible variation pattern of $A(T,\mu)$,
$B(T,\mu)$ and $C(T,\mu)$. We suppose at finite temperature and
zero chemical potential, the absolute value of $B(T,\mu)$ keeps
decreasing and tends to zero with temperature increasing, while
$A(T,\mu)$ first decreases to a negative value and then increases
with temperature increasing. $C(T,\mu)$ keeps positive in all the
cases. By this kind of variation, from Fig.\ref{f5}, one could see
that the system will evolve from the confined phase to the
deconfined phase across the axis origin, and the transition will
be second order. Thus we make the following ansatz of $A(T,\mu)$,
$B(T,\mu)$ and $C(T,\mu)$ as \bea
A(T,\mu)&=&a\left[(T-T_c)(k_1T-\0{1}{T_c})+\l_1\mu^2\right], \label{a}\\
B(T,\mu)&=&bExp\left[-k_2\left(\0{T+T_c}{T_c}\right)^6+k_2+\l_2\mu\right],
\label{b}
\\ C(T,\mu)&=&c,
\eea where $a, b$ and $c$ are the parameters of the FL model which
have been already given in section II. $k_1=4fm^2, k_2=0.15,
\l_1=0.5fm^2$ and $\l_2=0.1fm$ are the effective parameters of the
ansatz. $T_c$ is the critical temperature of the transition at
zero chemical potential which could be seen in later analysis. It
also serves as a temperature scaling factor which value can be
taken as $T_c=180MeV$. When $T=\mu=0$, it is clear that
$A(0,0)=a$, $B(0,0)=b$ and $C(0,0)=c$. One should notice that in
our ansatz with the temperature increasing the parameter
$B(T,\mu)$ will be infinitely close to zero but not zero. However
when the second order phase transition takes place, the absolute
value of $B(T,\mu)$ will be sufficiently small. At zero chemical
potential, from equation (\ref{a}), one could see at $T=T_c$,
$A(T_c,0)=0$. At the same time $B(T_c,0)\approx 0$. Thus the
deconfinement phase transition at zero chemical potential and
finite temperature takes place at $T=T_c=180MeV$ and the
transition order is second order. At zero temperature, from
equation (\ref{b}), one could see that $B(0,\mu)$ will never be
zero with chemical potential increasing, which means the
transition will be first order at zero temperature and finite
chemical potential.

\begin{figure}[tbh]
\begin{center}
\includegraphics[width=210pt,height=150pt]{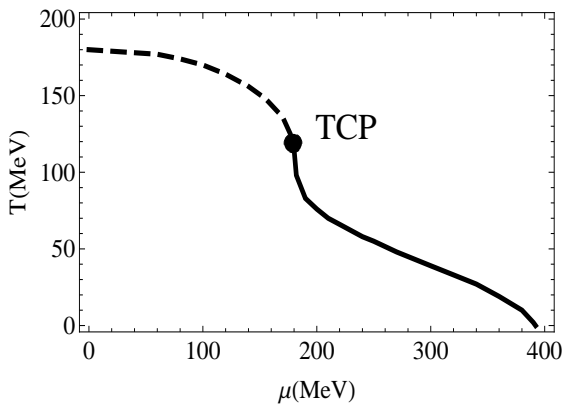}
\end{center}
\caption{$\mu-T$ phase diagram of deconfinement with TCP in the FL
model.}\label{f8}
\end{figure}

We can also evaluate the thermodynamic potential for different
chemical potentials and temperatures. At finite temperature and
zero chemical potential, the thermodynamic potential as a function
of $\bar\s$ is plotted in Fig.\ref{f6}. It is clear that the phase
transition is second order. At zero temperature and finite
chemical potential, it could be seen from Fig.\ref{f7} that the
transition is first order. The deconfinement phase transition
could be presented in a $\mu-T$ phase diagram as shown in
Fig.\ref{f8}. From first order phase transition to second order
phase transition there exists a TCP. The phase diagram is
qualitatively consistent with that of the chiral phase transition.
However, how to obtain the credible phase diagram of deconfinement
through the direct calculations including the fluctuations from
the Lagrangian of the FL model deserves a further investigation.

\section{summary}
In this paper we have discussed the possible phase diagram of
deconfinement in FL model. By the calculation only in the MFT
approximation and without the fluctuations, the deconfinement
phase transition can only be first order at finite temperature and
chemical potential. By the Landau expansion of the thermodynamic
potential and the analysis through Landau theory, we show that the
deconfinement phase transition can also be second order, which
will not appear in the MFT approximation but will possibly appear
when nonlinear fluctuations are considered. Thinking of the
difficulties in calculating the fluctuations, we have not done the
calculation here but made the ansatz that the Landau coefficients
are certain functions of temperature and chemical potential. By
this ansatz we obtain the possible $\mu-T$ phase diagram of
deconfinement in FL model which is similar to that of the chiral
phase transition. That means the deconfinement phase transition is
first order at low temperature and high chemical potential while
second order at high temperature and low chemical potential. From
first order to second order phase transition there exists a TCP.

\begin{acknowledgments}
This work was supported in part by the National Natural Science
Foundation of China with No. 10905018 and No. 10875050.
\end{acknowledgments}


\begin{thebibliography}{}
\bibitem{ref1}
M.~Gyulassy and L.~McLerran, Nucl. Phys. A750, (2005) 30.
\bibitem{ref2}
J.I.~Kapusta, J. Phys. G34 (2007) S295-304.
\bibitem{ref3}
B.~Svetisky and L.G.~Yaffe, Nucl. Phys. B210 (1982) 423.
\bibitem{ref4}
R.D.~Pisarski and F.~Wilczek, Phys.Rev. D29 (1984) 338.
\bibitem{ref5}
E.~Shuryak and T.~Schaefer, Phys. Rev. Lett. 75 (1995) 1707.
\bibitem{ref6a}
M.~Alford, K.~Rajagopal, and F.~Wilczek, Phys. Lett. B 422 (1998)
247; J.~Berges and K.~Rajagopal, Nucl. Phys. B 538 (1999) 215.
\bibitem{ref6b}
L.~McLerran and R.D.~Pisarski, Nucl. Phys. A796 (2007) 83-100;
L.~McLerran, K.~Redlich and C.~Sasaki, Nucl. Phys. A824 (2009)
86-100.
\bibitem{ref6c}
K.~Fukushima, Phys. Rev. D68 (2003) 045004;  Y.~Nishida,
K.~Fukushima and T.~Hatsuda, Phys. Rept. 398 (2004) 281-300.
\bibitem{ref6d}
F.~Karsch, Lect. Notes Phys. 583 (2002) 209-249.
\bibitem{ref7}
O.~Scavenius, A.~Mocsy, I.N.~Mishustin and D.H. Rischke, Phys.
Rev. C64 (2001) 045202.
\bibitem{ref8}
M.~Stephanov, K.~Rajagopal and E.~Shuryak, Phys.Rev.Lett. 81
(1998) 4816-4819.
\bibitem{ref9}
M.~Alford, K.~Rajagopal and F.~Wilczek, Phys.Lett. B422 (1998)
247-256.
\bibitem{ref10}
A.M.~Polyakov, Phys. Lett. B72 (1978) 477.
\bibitem{ref11}
B.~Svetitsky, Phys. Rept. 132 (1986) 1.
\bibitem{ref12}
K.~Fukushima, Phys. Lett. B591 (2004) 277.
\bibitem{ref13}
S.~Roessner, C.~Ratti and W.~Weise, Phys. Rev. D75 (2007) 034007.
\bibitem{ref14}
B.J.~Schaefer, J.M.~Pawlowski and J.~Wambach, Phys. Rev. D76
(2007) 074023.
\bibitem{ref15}
T.~Kahara and K.~Tuominen, Phys. Rev. D78 (2008) 034015.
\bibitem{ref16}
H.~Mao, J.~Jin and M.~Huang, J. Phys. G37 (2010) 035001.
\bibitem{ref17}
R.~Friedberg and T.D.~Lee, Phys. Rev. D15, (1977) 1694; D16,
(1977) 1096; D18, (1978) 2623.
\bibitem{ref18}
R.~Goldflam and L.~Wilets, Phys. Rev. D25 (1982) 1951.
\bibitem{ref19}
M.C.~Birse, Prog. Part. Nucl. Phys. 25 (1990) 1.
\bibitem{ref20}
H.~Reinhardt, B.V.~Dang and H.~Schulz, Phys. Lett. B159 (1985)
161.
\bibitem{ref21}
M.~Li, M.C.~Birse and L.~Wilets, J.Phys. G13 (1987) 1.
\bibitem{ref22}
E.K.~Wang, J.R.~Li and L.S.~Liu, Phys. Rev. D41 (1990) 2288;
S.~Gao, E.K.~Wang and J.R.~Li, Phys. Rev. D46 (1992) 3211;
S.H.~Deng and J.R.~Li, Phys.Lett. B302 (1993) 279.
\bibitem{ref23}
H.~Mao, R.K.~Su and W.Q.~Zhao, Phys. Rev. C74 (2006) 055204;
H.~Mao, M.J.~Yao and W.Q.~Zhao, Phys. Rev. C77 (2008) 065205.
\bibitem{ref24}
S.~Shu and J.R.~Li, Phys. Rev. C82 (2010) 045203.

\end{thebibliography}
\end{document}